# Predicting the Presence of Internet Worms using Novelty Detection

E. MARAIS, T. MARWALA

*School of Electrical and Information Engineering, University of the Witwatersrand, Private Bag 3, WITS 2050, South Africa*

**Abstract**. Internet worms cause billions of dollars in damage yearly, affecting millions of users worldwide. For countermeasures to be deployed timeously, it is necessary to use an automated system to detect the spread of a worm. This paper discusses a method of determining the presence of a worm, based on routing information currently available from Internet routers. An autoencoder, which is a specialized type of neural network, was used to detect anomalies in normal routing behavior. The autoencoder was trained using information from a single router, and was able to detect both global instability caused by worms as well as localized routing instability.

## Introduction

The Internet has grown into a vast structure combining thousands of smaller interlinked networks (Hunt 2002). Control of the Internet is decentralised, with each of the smaller networks transferring information using various protocols and standards. The *Border Gateway Protocol* (BGP) is the current standard for exchanging routing information on the Internet and is currently in its fourth version (Rekhter and Li 1995). BGP allows for each of the Internet's smaller networks to administer their own internal routing policies, and for arbitrary interconnections between these topologies to be specified (Paxson 1997). The decentralised topology and the use of BGP allow the Internet to scale effectively. Redundancy occurs due to the large number of interconnected routes. This redundancy, although providing a high level of failover caused by events such as router outages, has been shown to perform non-optimal path selection for routes (Savage, Collins et al. 1999).

The ease of access and widespread use of the Internet has made it open to attacks by worms which can be propagated through the Internet (Zou, Gong et al. 2002; Chen and Robert 2004). Worms are capable of spreading rapidly, such as in the case of the SQL Slammer worm which infected approximately 75,000 hosts in under ten minutes (Moore, Paxson et al. 2003). The worms' rapid propagation is able to cause wide-scale Internet routing instability. Routing instability is defined as the rapid change in network availability and topology information caused by various factors such as router configuration errors and physical and data-link problems (Labovitz, Malan et al. 1998).

The high load caused by certain worms has a documented effect in increasing the number of BGP Update messages sent between routers (Cowie, Ogielski et al. 2001; Dubrawsky 2003). The increase in traffic causes BGP sessions between routers to expire or fail due to BGP messages not received timeously or router failure. Routers are able to select alternative network paths and inform adjacent routers about the new routes, as well as about the loss of the previous route. This has the side-effect of increasing the number of BGP messages when a worm is present, which is the basis for the prediction in this research.

The study in (Moore, Shannon et al. 2003) lists the methods required for containing a worm. The first, *Reaction Time*, is the time taken to detect the worm, for which it is suggested that the method be automated to be effective. The second is the *Containment Strategy*, which is the means to isolate a computer from susceptible hosts. The last method is the *Blocking Location*, which is where the countermeasure system is deployed. The research for this study addresses the first method, providing an automated means to detect a worm as quickly as possible. The time taken to respond to a worm is currently measured in human time, often days before a fix is available (Moore, Shannon et al. 2003). This study proposes a technique to assist in automatically detecting the presence of a worm, using the information currently available from Internet routers.

## Routing on the Internet

Routing information provides details about the path for which data must be sent between a source and destination. This information is exchanged between the networks that form the Internet, as well as internally within these smaller networks. The smaller networks are able to manage their own internal routing using a variety of routing protocols, but it is necessary to use a common protocol between the networks of the Internet. The Border Gateway Protocol (BGP) has become the standard protocol for exchanging routing information on the Internet (Wang 2002). BGP routing transmits data between *Autonomous Systems* (AS), which are made up of a network or group of networks that have a common administration and routing policies (Y. Rekhter 1995). At a broad level, BGP is able to transmit routing information between many AS's, and is able to provide a high level of control and flexibility for inter-domain routing while still successfully managing policy and performance criteria. The use of AS's allow BGP devices to store only the routes between AS's, rather than the entire set of paths between all hosts on the Internet.

BGP routers exchange lists of advertised routing paths, informing neighboring routers about routes that are currently being used by a router for transmitting data. Each path exchanged between the routers includes the full list of hops between the routers, eliminating the possibility of loops. BGP routers exchange their full set of routing paths on initiation of a session, after which incremental updates are sent to inform of any changes. The protocol does not rely on any scheduled updates.

The BGP protocol runs on top of the TCP protocol (Rekhter and Li 1995), which already provides fragmentation, retransmission, acknowledgement and sequencing. TCP is already supported by BGP routers since this is the most used transmission protocol for delivering data on the Internet.

There are four types of messages supported by the BGP-4 protocol (Rekhter and Li 1995; Cowie, Ogielski et al. 2001):

- The *Open* message is the first command sent between BGP devices and is used to establish a BGP session. Once the session has been established, the devices are able to send other BGP messages.
- The *Update* message provides the latest information for the advertised routes used by the BGP device. It is used to simultaneously advertise new routes (known as *announcements*), and to inform of routes that have become unavailable (known as *withdrawals*). Advertised routes are those routes that the device is actively using for transmitting data, and does not include routing paths that it is aware of but not using.
- *Notification* messages are sent when an error forces the closure of the BGP session.
- The *Keep-Alive* message is transmitted between BGP devices to provide notification that the devices are still active. This is sent at timed intervals to ensure that the BGP session doesn't expire.

Two specific features of BGP are of specific interest to this project – the *Minimum Route Advertisement Interval* (MRAI) property and *BGP route flap damping*. MRAI is part of the original BGP protocol, and is implemented on the sender side to limit the time between successive BGP Update messages that provide information about the same AS (Rekhter and Li 1995). This is designed to limit route exploration (Labovitz, Ahuja et al. 2000), dealing with route instability in a time scale of tens of seconds (Chen 2000).

*BGP route flap damping* was introduced in the mid 1990's to reduce the effect of localised edge instability propagating outwards through the Internet (Chen 2000). It caters for instability on a longer time scale than MRAI, by reducing the effect of *route flapping*, which occurs when route availability constantly changes state. These state changes cause a router to continuously send frequent updates to neighbouring routers, potentially indicating localised instability. The research performed by Lad and colleagues indicates that BGP route flap damping has not been fully deployed across the Internet, but even if fully deployed the convergence time taken to recover from instability would be extended by route flap damping (Lad, Zhao et al. 2003).

## Worms and their Effect on BGP

A worm is a stand-alone automated program which exploits a network to infect other computers with copies of themselves (Chen and Robert 2004). A virus, on the other hand, is parasitically linked to a host program, and executed at the same time as the host program. Worms have become more prevalent as the Internet has expanded and more available bandwidth has become available, and it is expected that worms will remain a problem for the foreseeable future.

Although it is probably not the intention of the writers of worms to slow down the Internet, since this slows the rate at which worms spread, worms do have a significant effect on the stability of the Internet as a whole. The Code-Red worm first occurred during July 2001 and infected almost 360,000 computers in approximately 14 hours, reaching a peak of 2,000 infections per minute (Moore, Shannon et al. 2002), and caused an estimated $2.6 billion of damage. The Nimda worm appeared later in July 2001, and infected more than 450,000 hosts within 12 hours (Chen and Robert 2004). In the case of the SQL Slammer which appeared in January 2003, the worm infected 90 percent of susceptible hosts in ten minutes (Moore, Paxson et al. 2003). The cost of the worms is difficult to estimate, but includes time to analyse and create countermeasures, as well as the economic impact caused by the Internet downtime.

Worms have the ability to not only affect the computers at the endpoints of the Internet, but are also able to impact the infrastructure connecting these computers together. Dubrawsky showed the *Code-Red* and *Nimda* worms were responsible for the global Internet outages during 2001 (Dubrawsky 2003). Of particular interest to this project, it was also evident that the number of BGP Update messages increased significantly during these times. Edge network administrators reported "insane ARP storms", router failures, congestion slowdowns and connectivity problems during the Nimda attack (Cowie, Ogielski et al. 2001).

The increase in the number of BGP Update messages is referred to as "route flap storms", which are caused when routers fail during instability (Labovitz, Malan et al. 1998). The BGP session failure on routers can be caused by high CPU loads, exceeding available memory as well as cache overflows (Cowie, Ogielski et al. 2001). In the event of BGP session failure, the router's peers remove it from any AS paths in which it is used. This in turn forces these peer routers to select alternate routes that exclude the unavailable router, and in doing so several BGP announcement messages are sent from these routers. When the router either reboots or recovers from the instability, it attempts to re-establish connections with the peer routers, which in turn revert to using it again in their AS paths. This causes more BGP withdrawal messages to be sent. This is the scenario that the *BGP route flap damping* has been designed to assist with.

## Autoencoders for Novelty Detection

Novelty detection is the ability to distinguish a level of novel behaviour from what is considered normal behaviour (Thompson, Robert et al. 2002). This approach requires only normal behaviour to be defined, from which abnormalities can then be identified against this description (Tarassenko, Nairac et al. 2000). Systems trained to detect novelty are unable to detect faults, but can robustly identify anomalies even when the system has never seen a fault before.

Neural networks are suited to applications where the problem knowledge is not well understood and the objective function is unknown (Kasabov 1998). The auto-associative neural network (or *autoencoder*) is a specialised type of neural network that has the ability to detect novel behaviour. The network is trained using only normal behaviour, so that when data dissimilar to the training data is presented, the network is able to show this novelty. The novel data is not necessarily "good" or "bad", it simply falls outside of the range of data that the network has been trained with (Thompson, Robert et al. 2002).
The basic structure of an autoencoder consists of the same number of inputs as outputs. Training requires teaching the network to attempt to match each output to the corresponding input, so that the error between inputs and outputs is low for the normal or training data, but varies when the input dataset is unrecognised.

The autoencoder has the ability to perform information compression, recovery of missing sensory data and to implicitly learn characteristics of the underlying data (Thompson, Robert et al. 2002). Another feature of the autoencoder is to perform Principal Component Analysis, which is a widely-recognised method for dimensionality reduction, applications include pattern recognition and image processing (Tripping and Bishop 1997). The basic structure of an autoencoder is shown in Figure 1. The auto-associative nature of the network is provided in the training phase, with the same values used for inputs and outputs thereby ensuring that the error between input and output is minimal for non-novel datasets. The hidden layer provides a "bottleneck", mapping the input space onto a reduced dimensional space. The output of a trained network will be similar to the inputs if the input is similar to the training dataset. If the data provided to the trained network is not similar to the training dataset, the outputs will differ from the inputs. The autoencoder has previously been tested using both simulated and real-world data (Thompson, Robert et al. 2002), with promising results.

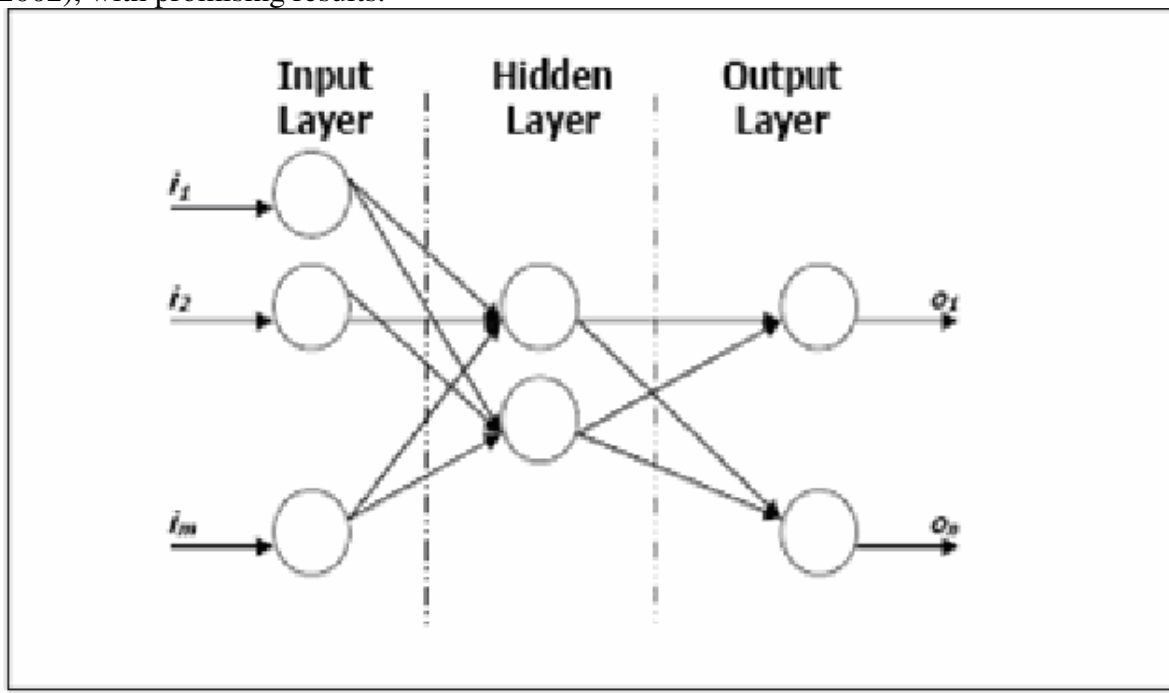

**Figure 1.** Basic autoencoder structure with the outputs trained to be as close to the inputs as possible for non-novel input values.

An alternative method to using a neural network is a *Symbolic AI rule-based system*, which is applicable for well defined problems with fixed rules, and when the variation of a more advanced system is not required or too difficult to construct (Kasabov 1998). These systems typically consist of several logical rules which define the behaviour of the system to various inputs.
This paper demonstrates a method of detecting the presence of a worm on the Internet, using data made available from a single Internet router. The type of system used for the prediction is an autoencoder (Thompson, Robert et al. 2002), which is a specialized type of neural network that provides a measure of novelty. The autoencoder is initially trained using only "normal" data, which ensures that the system is not trained to recognize only certain types of worms. This ensures that the worm is able to predict the presence of new worms, which are continuously released to exploit new vulnerabilities in software.

## Methodology

The data for this project was taken from the RIPE Network Coordination Centre project, which provides BGP routing data for several routing devices throughout the world. The specific router selected was the *rrc00.ripe.net*, which is based in Amsterdam, The Netherlands and collects data from 15 peer routers. The time range for the extracted data was between June and September 2001, during which the various Code-Red and Nimda worms affected Internet stability. The data consists of BGP routing information collected from the router from which only the BGP Update information was extracted and grouped into one-minute time intervals. These messages were separated into the number of BGP route announcement messages and withdrawal messages. The graph in Figure 2 displays the total number of BGP Update messages for the entire dataset.

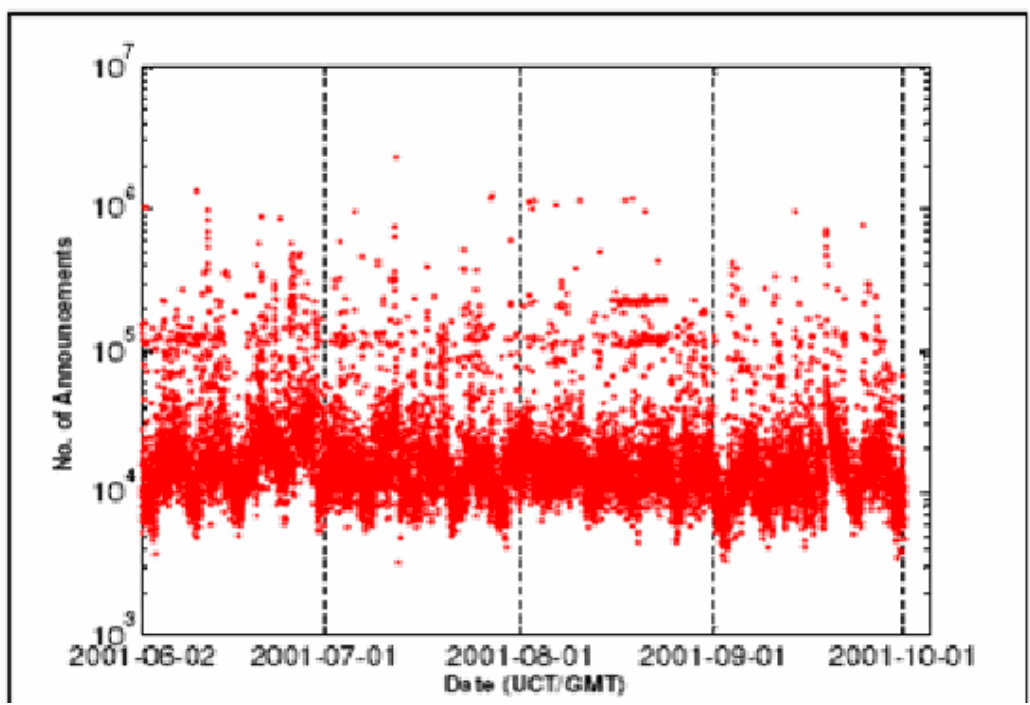

**Figure 2.** Logarithmic representation of the total number of BGP Updates for all the datasets between June and September 2001. These values were obtained from the *rrc0* router data used for this project.

A neural network was selected to provide the basis for the prediction, since the problem knowledge and objective function are not well understood. The autoencoder was specifically chosen for this problem, due to concerns that a neural network classifier would treat the increases in the number of BGP Updates during a worm as noise. The autoencoder also has the advantage of only requiring "normal" data for training, and is therefore not trained to recognise specific worms. A threshold value is specified from the novelty output of the autoencoder to determine when the novelty reaches warning levels, which would trigger an alarm for the presence of a worm in a real system.
The inputs to the autoencoder were the number of BGP announcements and withdrawals (which together add up to the number of BGP Update messages) within one-minute time intervals. Each dataset included the most-recent number of BGP announcements and withdrawals, as well as for several previous time periods, for which the prediction is required, allowing the autoencoder to learn underlying trends from the data.

The autoencoder provides several output values (one for each input), from which a single novelty value is calculated using the sum of the mean squares of the outputs:

$$e = \frac{1}{n}\sum_{1}^{n}\left((t_n - i_n)^2\right) \quad (1)$$

where $n$ is the number of inputs and outputs, $t_n$ is the $n^{th}$ output and $i_n$ is the $n^{th}$ input. The resultant error value $e$ is the overall level of novelty of the combined outputs.

The autoencoder was trained using data from the $2^{nd}$ to the $9^{th}$ of June, which was selected due to no known global Internet instabilities occurring for this time period. The training data was pre-processed using linear normalisation in the range [0, 1]. The neural network was trained using the *Scaled Conjugate Gradient* optimization technique, using 100 training cycles.

The primary goal for adjusting the various autoencoder parameters was to provide as distinct an indication of novelty as possible. This means that the autoencoder value should ideally provide either a near-zero (indicating no instability), or high value (indicating a possible worm).

In order to provide a comparison to a rule-based system, the data has been analysed to provide an indication of the times at which a rule-based system would trigger an alarm. A single rule is used to trigger an alarm of a possible worm, specifying that when the total number of BGP Updates exceeds a specified threshold, the alarm is triggered. The complexity from the point of view of a system administrator configuring this system is similar to that of the autoencoder, requiring that the threshold be determined in both cases. The time interval to group the BGP Update data for the rule-based system input is also one minute, the same as that used as input to the autoencoder.

## Results

The dates relevant to Internet stability, such as the appearance of worms and physical events that impacted the Internet are shown in Table 1.

**Table 1.** Dates of interest in terms of global Internet instability from June 2001 to September 2001. These dates include both worms and significant physical occurrences that affected the Internet.

| Date | Description |
| --- | --- |
| 13 July 2001 | Code-Red I v1 worm outbreak starts. |
| 18 July 2001 | Baltimore Tunnel Train Wreck |
| 19 July 2001 | Code-Red I v2 worm outbreak starts. |
| 4 August 2001 | Code-Red II worm outbreak starts. |
| 11 September 2001 | World Trade Centre Attacks |
| 18 September 2001 | Nimda worm outbreak starts. |

From experimentation, the neural network parameters were set as follows:
- 100 input-output pairs (50 BGP announcements and 50 BGP withdrawals)
- 100 hidden neurons
- 1 minute time intervals

The datasets did not include some values on the $5^{th}$ of July 2001, from approximately 17:15 to 19:09, due to unavailability of data. For this time period the values were treated as zero for both announcements and withdrawals during this period. There is therefore an expected increase in instability during this period. No reported global Internet instability could be found for June 2001.

The novelty output for the autoencoder for the entire time period is shown in Figure 3. Each of the significant Internet stability events is indicated on the graph, and is discussed in the following section. A threshold has been added to this graph, to provide an indication of when a real-system would provide an alarm. The threshold is set to include the most significant peaks.

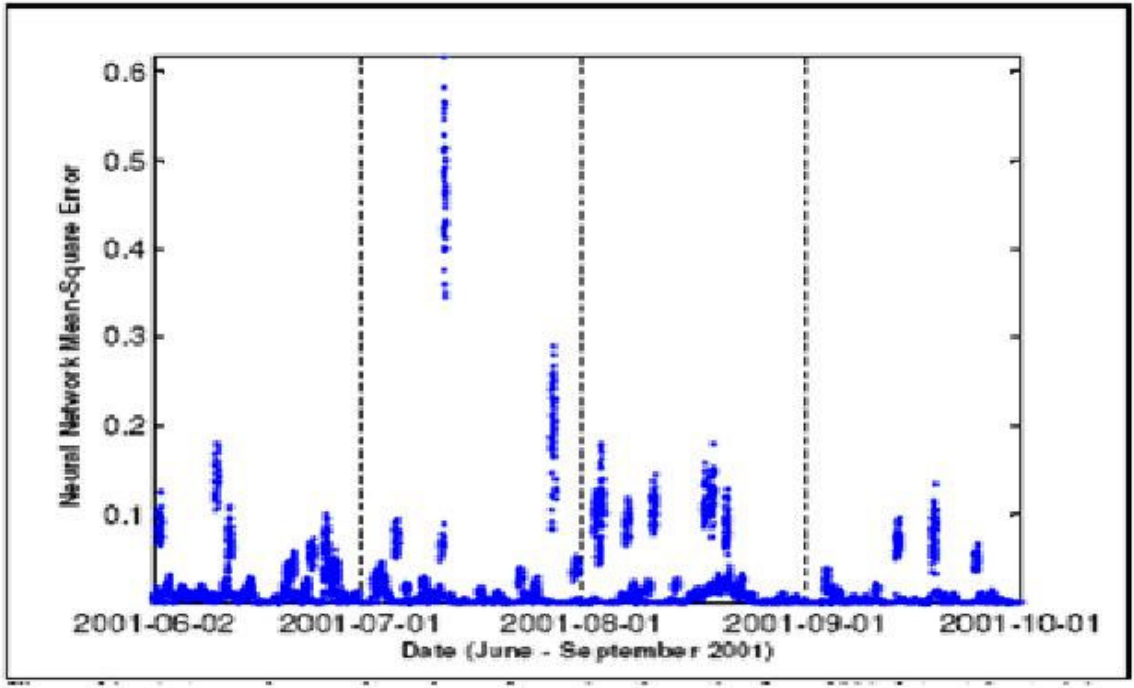

**Figure 3.** Novelty values from the final trained autoencoder for the entire dataset from June to September 2001. The network has been trained using 100 input-output pairs, 100 hidden neurons and the time interval for each point is one minute. A horizontal novelty threshold has been manually inserted.

In setting up the autoencoder, it was necessary to configure several parameters. It was found that modifying the number of input-output pairs, the number of hidden units and the size of the time interval did not clearly affect the novelty output. The choice of training data played a more significant role in the distinction of the peaks, but this also did not have a large impact on the novelty. By testing different training areas, the relative height of the different peaks was affected, but the same peaks were always evident. When including datasets that contained novelty spikes, which the autoencoder is then expected to treat as normal behaviour and so provide low novelty, the spikes were still evident although their relative height was smaller. This effectively demonstrates the autoencoder's ability to automatically perform Principle Component Analysis.

The top fifteen highest volume of BGP Updates is shown in Table 2, which provides an indication of the times that an alarm in a rule-based system would be triggered. The data is ordered from highest to lowest, and is for the entire dataset from June to September 2001. In order to give an indication of the effectiveness of this method, the table indicates the supposed stability event that each alarm was linked to.

It is noted that the increase in BGP Updates caused by the Nimda worm does not appear on this table – the highest increase caused by Nimda was the $25^{th}$ highest value. The autoencoder threshold was set to include the Nimda worm as the lowest novelty peak, and from Figure 3 it can be seen that there are twelve grouped peaks that exceed the threshold (seven of which are caused by the Code-Red II worm). By comparison, if the rule-based system threshold were set to include the Nimda worm, thirteen distinct alarms would have been triggered (grouped by similar times as shown in the table). This indicates that a similar number of alarms would have been triggered, using a similar threshold to the autoencoder, with each of the significant events corresponding between the rule-based system and the autoencoder.

The advantage of the autoencoder is evident from the expanded view of the novelty displayed in Figure 4 for the Nimda worm. This shows the novelty increasing before the high number of BGP Updates would have triggered an alarm in the rule-based system. The autoencoder shows a sharp increase in novelty about an hour before the rule-based system. This was not always the case with instability – the alarms for the Code-Red I v1 would have been triggered simultaneously for the autoencoder and rule-based systems.

**Table 2.** Highest number of BGP Updates for the entire dataset from June to September 2001. The time buckets are grouped into one-minute intervals, and the significant events noted.

| No. | Time (1min interval) | Number of BGP Updates | Event |
| --- | --- | --- | --- |
| 1 | 2001-07-27 14:50:00 | 595001 | Localised II (also seen with autoencoder). |
| 2 | 2001-08-02 13:30:00 | 592458 | Code-Red II |
| 3 | 2001-08-17 20:57:00 | 572124 | Code-Red II (extended effect) |
| 4 | 2001-08-18 20:39:00 | 556038 | Code-Red II (extended effect) |
| 5 | 2001-07-12 12:42:00 | 541756 | Code-Red I v1 |
| 6 | 2001-08-03 11:24:00 | 534271 | Code-Red II |
| 7 | 2001-08-20 20:44:00 | 526423 | Code-Red II (extended effect) |
| 8 | 2001-06-10 17:35:00 | 504463 | Localised I (also seen with autoencoder). |
| 9 | 2001-08-06 23:03:00 | 499349 | Code-Red II |
| 10 | 2001-06-10 17:36:00 | 486930 | Localised I |
| 11 | 2001-07-12 12:39:00 | 475865 | Code-Red I v1 |
| 12 | 2001-07-12 12:38:00 | 453161 | Code-Red I v1 |
| 13 | 2001-08-10 16:32:00 | 436326 | Code-Red II (extended effect) |
| 14 | 2001-08-10 16:33:00 | 432627 | Code-Red II (extended effect) |
| 15 | 2001-08-20 20:43:00 | 418252 | Code-Red II (extended effect) |

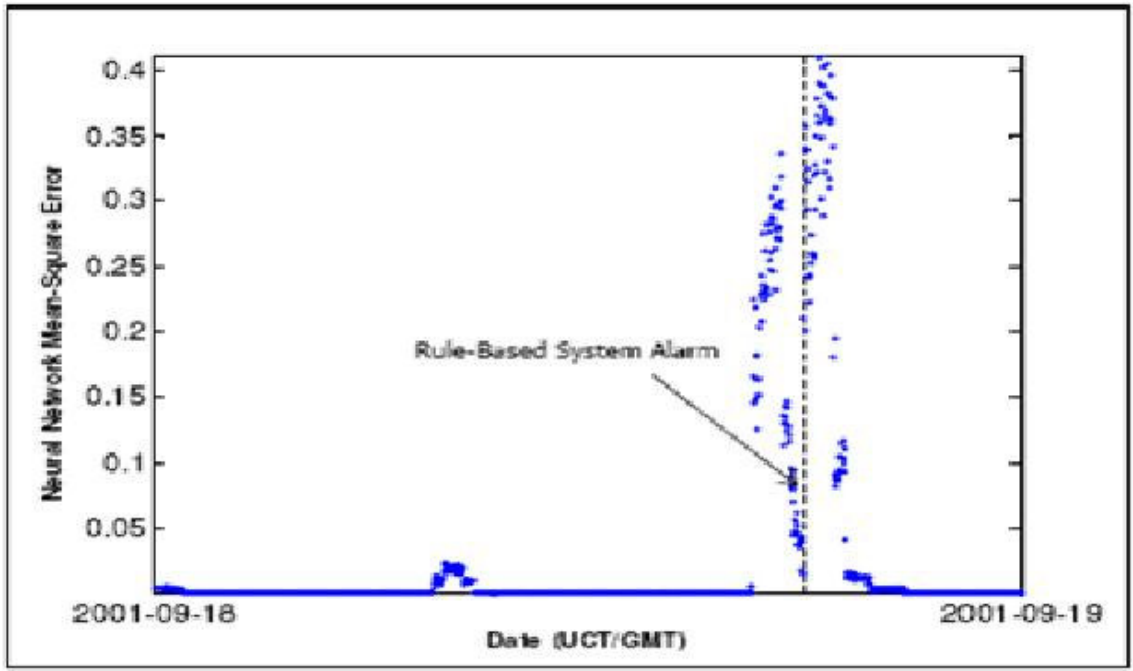

**Figure 4.** Comparison between autoencoder and rule-based system for Nimda. The graph shows the autoencoder output, and the grid line indicates the point at which the simple rule-based system would trigger an alarm.

## Discussion

The interpretation of the results seen in Figure 3 are discussed below:
- The first spike, *Localised I,* occurs during June 2001, which is not known to contain any global instability. The BGP routing data does contain a markedly higher number of Updates, which would trigger a rule-based alarm. This is suspected to be due to localised instability in the region of the *rrc0* router. The *Localised II* spike is again a suspected to be caused by localised routing instability within the neighborhood of the router, since there were no known occurences of a worm at that time.
- The second spike, *Zero Values*, is caused by the missing data described previously. The novelty values do not reach the threshold.
- The next spike corresponds to the *Code-Red I v1* worm, which caused global Internet instability and was expected to cause high novelty. This period contained 595,000 BGP announcements within a single minute, the highest number of BGP announcements for the entire dataset.
- The *Baltimore Tunnel Train Wreck* caused a very small increase in novelty. This incident triggered a localised problem in another part of the Internet that caused a burst of BGP Updates, after which the routing converged. The BGP protocol is able to deal with this type of sudden change due to the high level of redundancy in the Internet.
- The *Code Red I v2* worm showed a small increase in novelty, which was low since the number of BGP Updates at this time was not within the top 25 values. From the literature, this worm was designed to stop propagating on the 20th day of the month, which was the the day after the worm occurred. This therefore appears to have severly limited the impact of the worm.
- The appearance of *Code-Red II* corresponds with the spike in early August, with the peak well above the threshold, indicating a threat to stability. The novelty output remains high during August, most likely due to this worm, as well as the Code-Red I strains which have been shown to be active during this time.

The research performed by Moore showed that 32% of Code-Red I infected hosts had not been patched by 1 August 2001, the time at which the worm restarted its propagation phase (Moore, Shannon et al. 2002). It therefore seems likely that these spikes were due to the various Code-Red I and II versions.
- The increase caused by the *World Trade Centre Attacks* on 11 September 2001 are also caused by local instability. Similar to the Baltimore Tunnel Train Wreck, this localised instability causes short localised instability, after which paths stabilised.
- The last two spikes are suspected to be due to the Nimda worm, which appeared on 18 September 2001.

The results have confirmed that the number of BGP Update messages provide a good indication of instability, both localised within the neighborhood of the router being monitored, and globally by the presence of worms. The autoencoder is unable to distinguish between the neighboring localised instability and the global instability, but the occurence of localised instability in disparate areas of the Internet (such as the World Trade Centre Attack) increase the autoencoder novelty, but not sufficiently to trigger an alarm.

It is suspected that correlating the novelty received from autoencoders that are monitoring disparate portions of the Internet will indicate which of the spikes are caused by localised instability, leaving only global problems. A concern for using this technique is that correlation would require sending the data to a central point, most likely using the Internet for transmission, which could be problematic when a worm is present.

The autoencoder was not particularly sensitive to changes in the various neural network parameters, such as the number of input-output pairs or the number of hidden units. The implication of this resiliance to changes in parameters means that the configuration is straightforward, which would be useful for real implementation.

The autoencoder was also able to respond to a sharp increase in the number of routing messages quickly. In the comparison to the rule-based system, it was shown that in some cases the autoencoder was able to indicate a problem sooner than a simple rule-based system.

## Conclusion

Worms are able to disrupt the Internet, impacting world-wide productivity and the global economy. New worms are continuously released to take advantage of new vulnerabilites in software, and each new worm has a different signature. This study has assumed that worms will remain a problem, and addresses the issue of identifying the presence of a worm faster than human detection. The research extends from previous work, which showed that there is an increase in the number of BGP Update messages sent when there is global instability caused by a worm. The autoencoder was demonstrated to show a high novelty during global Internet instability, as well as localised instability within the vicinity of the router being monitored. Localised instability in disparate areas of the Internet, such as that caused by the World Trade Centre Attacks, did not increase the novelty to alarm levels. The autoencoder was shown to provide similar alarms based on the number of BGP Updates, but in some cases could provide warnings sooner than a rule-based system. The autoencoder provides an automated means of detection that is faster than that of a human, and has the potential to reduce the damage done by worms' propagation and payload. It provides a reactive warning of the presence of a worm, only after the worm has started to spread and cause global instability. This work is the first known application of a neural network to detect Internet instability using BGP routing information. An extension of this work would be to correlate the novelty values from multiple routers, to be able to distinguish local and global instability. The introduction of additional inputs would allow the autoencoder to learn more information about the system. An example of an additional input would be to use the number of distinct AS's for which BGP Updates were received within the time interval. A large number could provide an indication that the instability is global, since localised instability would be from a small number of AS's.

# Acknowledgements

We would like to thank ARMSCOR for the financial support, and the CSIR for the technical support and discussions.